\documentclass[twocolumn,preprintnumbers,amsmath,amssymb]{revtex4-1}
\usepackage{graphicx}
\usepackage{dcolumn}
\usepackage{amsmath,amsbsy,amssymb,scalefnt}
\usepackage{bm}
\usepackage{epstopdf}
\usepackage[T1]{fontenc}
\usepackage{amsfonts}
\usepackage{textcomp}
\usepackage{mathtools}
\usepackage{hyperref}
\hypersetup{backref=true,
 pdfnewwindow=true, colorlinks=true,
 linkcolor=blue, anchorcolor=blue,
 citecolor=blue, filecolor=blue,
 menucolor=blue, urlcolor=blue}

\begin{document}
\title{\large\bf Strain-induced topological phase transition in phosphorene and in phosphorene nanoribbons  }
\author{E. Taghizadeh Sisakht,$^{1,2}$ F. Fazileh,$^2$ M. H. Zare,$^3$ M. Zarenia,$^1$ and F. M. Peeters$^1$
 \small $^1$Department of Physics, University of Antwerp, Groenenborgerlaan 171, B-2020 Antwerpen, Belgium.\\
 \small $^2$Department of Physics, Isfahan University of Technology, Isfahan 84156-83111, Iran.\\
 \small $^3$Department of Physics, Faculty of Science, Qom University of Technology, Qom 37181-46645, Iran.
 }
 \date{\today}

\begin{abstract}

Using the tight-binding (TB) approximation with inclusion of the spin-orbit interaction, we predict a topological phase
transition in the electronic band structure of phosphorene in the presence of axial strains. We derive a low-energy TB Hamiltonian that includes the spin-orbit interaction
for bulk phosphorene. Applying a
compressive biaxial in-plane strain and perpendicular tensile strain in ranges where the structure is still stable,
leads to a topological phase transition. We aslo, examine the influence of strain on zigzag phosphorene nanoribbons (zPNRs) and the formation of the corresponding
protected edge states when the system is in the topological phase. For zPNRs  up to a width of $100$~nm  the energy gap is at least three orders of 
magnitude larger than the thermal energy at room temperature.

\vspace{5mm}
\vspace{5mm}
\end{abstract}
 \maketitle

\section{INTRODUCTION}
Topological insulators (TIs) with time-reversal symmetry (TRS), have been of increasing interest 
in condensed matter physics and material science during the last decade. The emergence  of  robust edge states in two-dimensional (2D) TIs that are protected by TRS, make them promising  candidates for 
potential applications in spintronics and quantum computing~\cite{kane,hasan,moor,qi,fu,bernevig}. TIs can exist intrinsically
or be driven by external factors such as electrical field or by functionalization~\cite{ren}. Strain engineering is a well known strategy for
switching from normal insulator (NI) phase to a TI phase~\cite{ren,ma}. Among the wide list of systems that possesses such property, 
2D materials with fascinating electronic, mechanical and thermal properties have been in the focus of attention~\cite{kane,ezawa1}.\par
In the past few years, phosphorene, a monolayer of black phosphorus, has emerged as an encouraging 2D semiconducting material 
for widespread applications. Phosphorene-based field effect transistors (FETs), show a higher ON/OFF ratio in comparison with graphene~\cite{koenig,li} and has a higher
carrier mobility with respect to 2D transition metal dichalcogenides (TMDs) which have recently attracted 
a lot of attention for FET applications~\cite{koenig,li,xia}. There exist several works pertinent to the observation of different phases in bulk and multilayer black phosphorous by 
tuning the lowest energy bands~\cite{fei1,kim,xiang,zunger,zhang}.
Using density functional theory (DFT) it was shown that few-layers of phosphorene experiences
a NI to TI and then a TI to topological metal (TM) phase transition by applying a perpendicular electric field~\cite{zunger}. In a different DFT study~\cite{zhang} such phase
transitions for various stacked bilayer phosphorene under in-plane strain has been explored.
Owing to the puckered structure of phosphorene, it has a high degree of flexibility. Therefore, it can sustain strain very well 
specially in the zigzag direction up to about 30\%~\cite{wei,peng}. This makes phosphorene promising for possible applications  using strain engineering.\par
In our work, we investigate the effect of strain on the electronic band structure of phosphorene the TB approach.
The band gaps of this model~\cite{rud} are close to the most reliable DFT and experimental results~\cite{liang,tran} that predict band gaps of $1\sim2$ eV for phosphorene.
In this paper, we propose a model Hamiltonian for the SOC 
for monolayer phosphorene that can be generalized to few-layers phosphorene. We show that, a model which includes the
next-nearest(n-n) neighbors  in the upper or lower chains, is sufficient for capturing the main physics. 
Then, strain engineering of this system is investigated through modifying the hopping parameters of the system. We demonstrate that, by applying particular types of strain, the system can make a phase transition to a TI. 
Finally, we show numerically that though the topological bulk band gaps induced 
by SOC is about $5$ meV, but the highly anisotropic nature of this material causes 
the corresponding bulk gaps in large widths 
zPNRs be at least three order of magnitude  larger than room temperature thermal energy ($\sim26$ meV) and makes 
phosphorene nanoribbons excellent candidates for future applications.

This paper is organized as follows: the effective low-energy TB model Hamiltonian including the SOC terms is obtained in Sec.~II. 
The effect of  axial strains on the band structure  produced by this model is calculated and our results are compared with DFT results in Sec.~III. Demonstration of a 
topological phase transition in the electronic properties of phosphorene when particular types of strain are applied  and the characteristics of corresponding 
edge states in zPNRs is presented in Sec.~IV. The paper is summarized in Sec.~V.
~~~~~~~~~~~~~~~~~~~~~~~~~
\begin{small}\section{TIGHT-BINDING MODEL HAMILTONIAN INCLUDING SPIN-ORBIT INTERACTION}

\begin{figure*}
\centering
{\includegraphics[angle=0,width=0.8\textwidth]{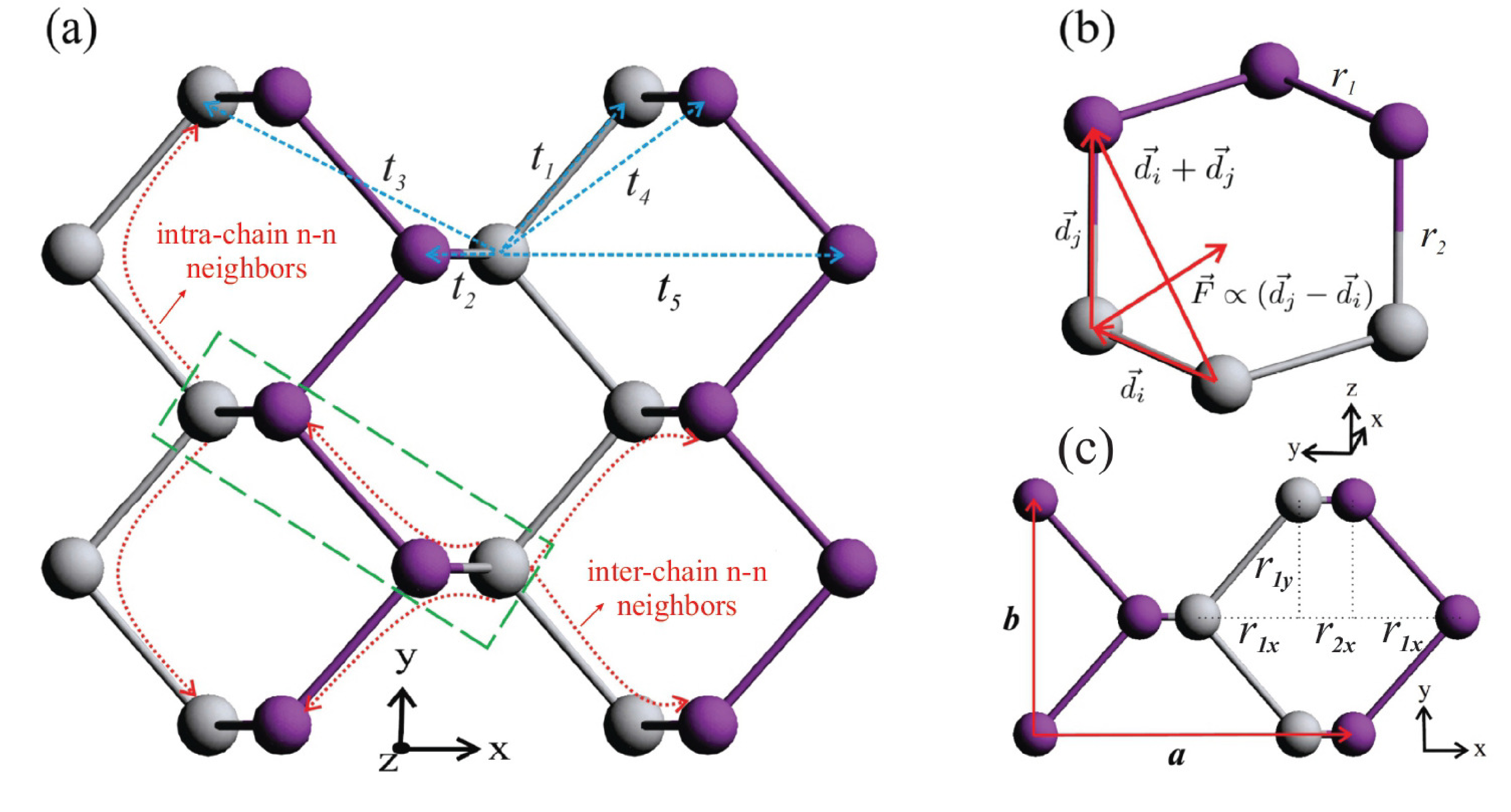}}
 \caption{ The lattice geometry of phosphorene. The two different colors of the P atoms refer to upper and lower chains (a) The hopping parameters $t_1,t_2,...,t_5$ used in our TB model are indicated in the figure. Red dotted  arrows represent two types of 
 n-n neighbors and the green dashed rectangle shows the unit cell of phosphorene. (b) A honeycomb-like ring of phosphorene. The
 vectors $\vec{d}_i$, $\vec{d}_j$, $\vec{d}_i+\vec{d}_j$ and $\vec{F}\propto(\vec{d}_j-\vec{d}_i)$  are used to derive the SOC.
 (c) Lattice constants and the components of geometrical parameters describing
 the structure of phosphorene.}
\label{lattice}
\end{figure*}
~~~~~~~~~~~~~~~~~~
\subsection{Structure}\end{small}
The puckered atomic structure of phosphorene and its geometrical parameters are shown in Fig.~\ref{lattice} where the $x$
and $y$ axes are the armchair and zigzag directions, respectively and the $z$ 
axis is in the normal direction to the plane of phosphorene. With this definition of coordinates, one can indicate the various atom connections $r_i$ 
which correspond
to various hopping parameters $t_i$  that are included in  the TB model.
The structure parameters have been taken from~\cite{qiao} which is
very close to experimentally measured parameters~\cite{takao} for its bulk structure. The components of the geometrical parameters as shown in 
Figs.~\ref{lattice}(b) and (c), for bond lengths 
$r_1=2.240$ \r{A} and $r_2=2.280$ \r{A}
are $(r_{1x},r_{1y},r_{1z})=(1.503,1.660,0)$ and $(r_{2x},r_{2y},r_{2z})=(0.786,0,2.140)$, and $r_3,r_4,r_5$ are
simply defined by parameters of $r_1$ and $r_2$. The two in-plane lattice constants are $a=4.580$ \r{A}, $b=3.320$ \r{A} 
and the thickness of a single layer due to the puckered nature is $r_{2z}=2.140$ \r{A}.

~~~~~~~~~~~~~~~~~~~~~~~~~
~~~~~~~~~~~~~~~~~~~~~~~~~~~
\subsection{Tight-binding model}
The phosphorene TB Hamiltonian that has been proposed earlier~\cite{rud}, without the spin degree of freedom, is given by
\begin{equation}
\hat{H}=\sum_{ i,j }t_{ij}c_{i}^{\dagger }c_{j},
\end{equation}
where the summation is up to fifths neighbors, and $t_{ij}$ are
hopping integrals that show the energy transfer between the $i$th and $j$th sites. The hopping terms are 
shown in Fig.~\ref{lattice}(a). $c_i^{\dagger}$ and $c_{j}$ represent the creation and annihilation operators of electrons in sites $i$ and $j$,
respectively. The numerical values of these hopping parameters are: $t_1=-1.220$~eV, $t_2=3.665$~eV, $t_3=-0.205$~eV, $t_4=-0.105$~eV,
and $t_5=-0.055$~eV~\cite{rud}. Including the spin degree of freedom and SOC the Hamiltonian is modified into
\begin{equation}
\hat{H}=\sum_{ i,j,\alpha }t_{ij}c_{i\alpha}^{\dagger }c_{j\alpha}+\hat{H}_{SO},
\label{H}
\end{equation}
where in $\hat{H}_{SO}=\hat{H}_{SO1}+\hat{H}_{SO2}$, the first term is called the usual effective SOC and the second one is the intrinsic Rashba SOC which will be introduced in next subsection. Due to the puckered 
structure of phosphorene, the Rashba term is rather large as compared to the first term and should be included in our calculations.

\subsection{Spin-orbit coupling in Phosphorene}
The primary goal of this subsection is to introduce a spin-orbit model Hamiltonian for phosphorene which  can capture the most important spin-related 
phenomenon. There exist several studies which showed the anisotropic
behaviour in the electronic and optical  properties of phosphorene~\cite{fei,yang,tran,sisakht} which are 
due to the anisotropic nature of the band dispersion of phosphorene. This property is reflected in the effective mass of electrons 
and holes of phosphorene. As a matter of fact, the corresponding band dispersion of the zigzag direction in real space, is relatively flat near the Fermi energy while it has an approximately
linear dispersion in  the armchair direction~\cite{fei,sisakht}. One can define  two types  of  n-n neighbors
in the phosphorene structure. As shown in Fig.~\ref{lattice}(a), each P atom  
has two intra-chain  and four  inter-chain n-n 
neighbors, respectively. The effective mass of electrons in the direction of intra-chain, are at least an order of  magnitude larger than the inter-chain  direction~\cite{fei}. Therefore, electrons usually 
select the inter-chain path for circular motion, allowing us to  ignore the intra-chain neighbors and
only consider the four n-n inter-chain P atoms in the SOC model.

In general, the SOC term for a 2D system is given by

\begin{equation}
H_{SO} =-\frac{\hbar}{4m_0^2c^2}(\vec{F}\times\vec{P})\cdot\vec{\sigma},
\end{equation}\\
where $\hbar$, $m_0$ and $c$ are Plank's constant, mass of free electron, and the velocity of light, respectively. $\vec{F}$ is the effective 
electrostatic force, $\vec{P}$ is the effective  momentum and $\vec{\sigma}$ denotes the Pauli matrices. As in the cases of graphene 
and silicene~\cite{liu}, the nearest-neighbor SOC is zero in phosphorene, but the SOC terms of the n-n 
neighbors are nonzero.

 As shown in Fig.~\ref{lattice}(b), in a honeycomb-like ring  of phosphorene, we can define $\vec{d}_i$ and $\vec{d}_j$ as vectors that
 connect the nearest P atoms to each other and $\vec{d}_i+\vec{d}_j$  the connecting
vector of n-n neighbors. Using these vectors, the electrostatic force and momentum can be written as  
$\vec{F}=|\vec{F}|(\vec{d}_j-\vec{d}_i)/|\vec{d}_j-\vec{d}_i|$ and 
$\vec{P}=-i\hbar\vec{\triangledown}\equiv-i\alpha(\vec{d}_i+\vec{d}_j)$, with $\alpha$ being a prefactor. Rewriting the SOC in terms of the above definitions we obtain

\begin{equation}
H_{SO} =-\frac{\hbar}{4m_0^2c^2}[   \frac{|\vec{F}|(-i\alpha)}{|\vec{d}_j-\vec{d}_i|}(\vec{d}_j-\vec{d}_i)\times(\vec{d}_i+\vec{d}_j) ]\cdot\vec{\sigma}.
\end{equation}\\
 Based on  experimental and DFT data,  $|\vec{d}_i|$ and $|\vec{d}_j|$ are approximately equal~\cite{qiao,takao,gomes,wei}, therefore $(\vec{d}_i+\vec{d}_j)$ and $(\vec{d}_j-\vec{d}_i)$ 
become perpendicular 
to each other. This leads to

\begin{equation}
\small H_{SO} =-i\frac{2\hbar\alpha|\vec{F}|}{4m_0^2c^2|\vec{d}_j-\vec{d}_i|}(\vec{d}_i\times\vec{d}_j)\cdot\vec{\sigma}\equiv-i\gamma(\vec{d}_i\times\vec{d}_j)\cdot\vec{\sigma},
\label{HSO}
\end{equation}%
where the term $2\hbar\alpha|\vec{F}|/4m_0^2c^2|\vec{d}_j-\vec{d}_i|=\gamma$ will be adjusted to obtain the correct value of SOC as obtained by DFT. Notice that, the 
above approximations reduce
the two parameters of the usual SOC and intrinsic Rashba SOC into a single parameter.
Using $\vec{\sigma}=\sigma_\shortparallel\hat{a}_\shortparallel+\sigma_z\hat{a}_z$, where
$\sigma_\shortparallel$ ($\sigma_z$) are the in-plane (out of plane) Pauli matrixes (matrix), we rewrite Eq.~(\ref{HSO}) as

\begin{equation}
\small H_{SO} =-i\gamma|\vec{d}_i\times\vec{d}_j|_z\nu_{ij}\sigma_z -i\gamma|(\vec{d}_i\times\vec{d}_j)_\shortparallel|(\vec{d}_i\times\vec{d}_j)_\shortparallel^0\cdot\vec{\sigma_\shortparallel},
\label{HSO1}
\end{equation}%
where $\nu_{ij}\equiv(\vec{d}_i\times\vec{d}_j)_z/|\vec{d}_i\times\vec{d}_j|_z=\pm1$ and
$(\vec{d}_i\times\vec{d}_j)_\shortparallel^0\equiv(\vec{d}_i\times\vec{d}_j)_\shortparallel/|(\vec{d}_i\times\vec{d}_j)_\shortparallel|$
 is a dimensionless unit vector. The spin-orbit terms in second quantization are given by

\begin{eqnarray}
\small \hat{H}_{SO1}+\hat{H}_{SO2}=-i\lambda_{so}\sum_{\ll ij\gg\alpha\beta}\nu_{ij}c_{i\alpha}^\dag\sigma_z^{\alpha \beta} c_{j\beta} \notag \\
-i\lambda_{r}\sum_{\ll ij\gg\alpha\beta}c_{i\alpha}^\dag(\vec{d}_i\times \vec{d}_j)^0\cdot\vec{\sigma}_\shortparallel^{\alpha \beta} c_{j\beta},
\label{HSO2}
\end{eqnarray}\\
where $\lambda_{so}\equiv\gamma|\vec{d}_i\times\vec{d}_j|_z$ and $\lambda_r\equiv\gamma|(\vec{d}_i\times\vec{d}_j)_\shortparallel|$ are effective intrinsic
SOC and intrinsic Rashba 
constants, and the summation runs over the 
inter-chain n-n neighbors.
As mentioned before, these two parameters are related to one parameter $\gamma$, which can be estimated by adjusting
the TB band structure of phosphorene to the one obtained from  DFT. It was shown that in the absence of SOC the energy gap of few-layers phosphorene closes under an external electric field or strain~\cite{zunger,zhang}.
However, including the SOC an energy gap of $5$~meV~\cite{zunger} remains in few-layers phosphorene. This results in the value of $\gamma\approx0.006$ meV/$\mathring{A^2}$ in our TB model.

\section{PHOSPHORENE UNDER STRAIN: ELECTRONIC BAND STRUCTURE}
The role of uniaxial and biaxial strain in manipulating the electronic structure of few-layers phosphorene has been investigated
via DFT~\cite{rodin,peng,wang,zhang,huang} and TB approaches~\cite{jiang,mohammadi,duan}. Applying tensile or compressive strain 
in different directions results in different modifications of the electronic bands. One can observe a direct to indirect gap transition, or a prior direct 
band gap closing, depending on the type of applied strain~\cite{peng,wang,zhang}.
In this work we consider biaxial compressive strain in the plane of few-layers phosphorene~\cite{wang,zhang}, and 
tensile strain in the normal 
direction~\cite{huang}. This modifies the low energy bands so that
the valance and conduction bands approach each other.
By further increasing strain, the lower band, coming from $p_x$ orbitals, shifts upward resulting in a
semi-metal phase~\cite{wang} given that at the band crossing point a mini gap opens due to the SOC. Investigating the local density of states of $p$ orbitals~\cite{zhang} shows that our 
one orbital $p_z$-like TB model is still valid in the low energy limit before the semi-metal phase appears.

In the following, we will first study the bulk band of phosphorene in the presence of  axial strains using our TB approach and demonstrate that a band inversion occurs in the energy spectrum of phosphorene in the range where the 
structure is still stable under strain. It has been shown that the bond lengths and bond angles  of phosphorene both change
under axial strains~\cite{wang,sa}. Therefore, the hopping parameters will change. According to the Harisson  rule~\cite{harrison,tang}, the hopping parameters for $p$ orbitals are  related to the bond length
as $t_i\propto 1/r_i^2$ and the angular dependence can be described by the hopping integrals along
the $\pi$ and $\sigma$  bonds. However, our calculations showed that, though the changes in angles are almost noticeable~\cite{wang,sa}, the modification of
the hopping parameters due to them is much smaller than the effect of changes of bond lengths. Hence, 
we consider only changes of the bond lengths in the hopping modulation.

When an axial strain is applied to phosphorene, the rectangle shape of the unit cell with lattice constants of $a_0$ and $b_0$ remains unchanged. Therefore the initial geometrical parameter $r^0_i$
is deformed 
as $(r_{ix}, r_{iy}, r_{iz})=((1+\varepsilon_x)r^0_{ix}, (1+\varepsilon_y)r^0_{iy}, (1+\varepsilon_z)r^0_{iz})$
where $\varepsilon_j$ is the strain in the $j$-direction and $r_i$ is a deformed geometrical parameter. In the linear 
deformation 
regime, expanding the
norm of $r_i$ to first order of $\varepsilon_j$ gives
 
\begin{equation}
r_{i}=(1+\alpha_x^i\varepsilon_x+\alpha_y^i\varepsilon_y+\alpha_z^i\varepsilon_z)r^0_i,
\end{equation}
where $\alpha_j^i=(r^0_{ij}/r^0_i)^2$ are coefficients related to the structure of phosphorene which are
simply calculated via the special geometrical parameters given in previous section. Using the Harrison relation,
we obtain the strain effect on the hopping parameters as
\begin{equation}
t_i\approx(1-2\alpha_x^i\varepsilon_x-2\alpha_y^i\varepsilon_y-2\alpha_z^i\varepsilon_z)t^0_i,
\label{MODIFIED_T}
\end{equation}
where $t_i$ is the modified hopping parameter of deformed phosphorene with new lattice constants $a$ and $b$.

Let us now study the energy spectrum of strained phosphorene with the modified hopping parameters as given by Eq.~(\ref{MODIFIED_T}). The unit 
cell of monolayer phosphorene is a rectangle containing four atoms as shown in Fig.~\ref{lattice}(a).
Fourier transform of the  strained Hamiltonian of Eq.~(\ref{H}) gives the 
general Hamiltonian in momentum space as

\begin{equation}
H=\sum_{\bf k} \psi^{\dagger}_{\bf k}H_{\bf k}\psi_{\bf k},
\label{Hk1}
\end{equation}
where we have used the basis  $\psi^{\dagger}_{\bf k}=\{a^{\dagger}_{\bf k},b^{\dagger}_{\bf k},c^{\dagger}_{\bf k},d^{\dagger}_{\bf k}\}\otimes{\{\uparrow,\downarrow\}}$
with $H_{\bf k}$ being 

\begin{equation}
H_{\bf k} = \begin{pmatrix}  H_{\bf k}(\uparrow\uparrow)&  H_{\bf k}(\uparrow\downarrow) \\
H_{\bf k}(\downarrow\uparrow) &  H_{\bf k}(\downarrow\downarrow)  
\label{Hk2}\end{pmatrix},
\end{equation}\\
where

\begin{eqnarray}
H_{\bf k}(\uparrow\uparrow)&=&H^{(4)}_{\bf k}+H^{so}_{\bf k},~~H_{\bf k}(\downarrow\downarrow)=H^{(4)}_{\bf k}-H^{(so)}_{\bf k},  \notag \\
H_{\bf k}(\uparrow\downarrow)&=&H^{(R)}_{\bf k},~~~~~~~~~~~~~H_{\bf k}(\downarrow\uparrow)={H^{\dagger}_{\bf k}}^{(R)},
\end{eqnarray}\\
are $4\times4$ matrices

\begin{eqnarray}
H^{(4)}_{\bf k} &=&\begin{pmatrix} 0 & A_{\bf k}& B_{\bf k} & C_{\bf k}  \\
A^{*}_{\bf k} & 0 & D_{\bf k} & B_{\bf k}  \\ 
B^{*}_{\bf k} & D^{*}_{\bf k} & 0 & A_{\bf k}  \\
C^*_{\bf k} &B^{*}_{\bf k} & A^{*}_{\bf k} & 0  
\end{pmatrix} , \notag \\
H^{(so)}_{\bf k} &=& \begin{pmatrix} 0 & 0 & E_{\bf k}& 0  \\
0 & 0 & 0 & -E_{\bf k}  \\ 
{E^{*}_{\bf k}} & 0 & 0 & 0  \\
0 & -{E^{*}_{\bf k}} & 0 & 0  
\end{pmatrix} ,\notag \\
H^{(R)}_{\bf k} &=& \begin{pmatrix} 0 & 0 & F_{\bf k}& 0  \\
0 & 0 & 0 & F_{\bf k}  \\ 
{e^{i(k_a-k_b)}F_{\bf k}} & 0 & 0 & 0  \\
0 & {e^{i(k_a-k_b)}F_{\bf k}} & 0 & 0  
\label{Hk3}
\end{pmatrix},
\end{eqnarray}

whose elements are given by

\begin{eqnarray}
\small A_{\bf k} &=& t_2+t_5e^{-ik_a},\notag \\
\small B_{\bf k} &=& 4t_4e^{-i(k_a-k_b)/2}\cos(k_a/2)\cos(k_b/2),\notag \\
\small C_{\bf k} &=& 2e^{ik_b/2}\cos(k_b/2)(t_1e^{-ik_a}+t_3), \notag \\
\small D_{\bf k} &=& 2e^{ik_b/2}\cos(k_b/2)(t_1+t_3e^{-ik_a}), \notag \\
\small E_{\bf K} &=& -2\lambda_{so}e^{-i(k_a-k_b)/2}\sin(k_a/2)\sin(k_b/2), \notag \\
\small F_{\bf K} &=& 4\lambda_re^{(k_b-k_a)/2}(\cos(k_b/2)\cos(k_a/2)\cos(\theta), \notag \\
&&+i\sin(k_b)\sin(k_a)\sin(\theta)),\notag \\
\end{eqnarray}
 with $k_a=\textbf{k.a}$, $k_b=\textbf{k.b}$ and $\theta=\arctan(r_{1y}/r_{1x})$.\\
 
\begin{figure}
\centering
{\includegraphics[angle=0,width=0.48\textwidth]{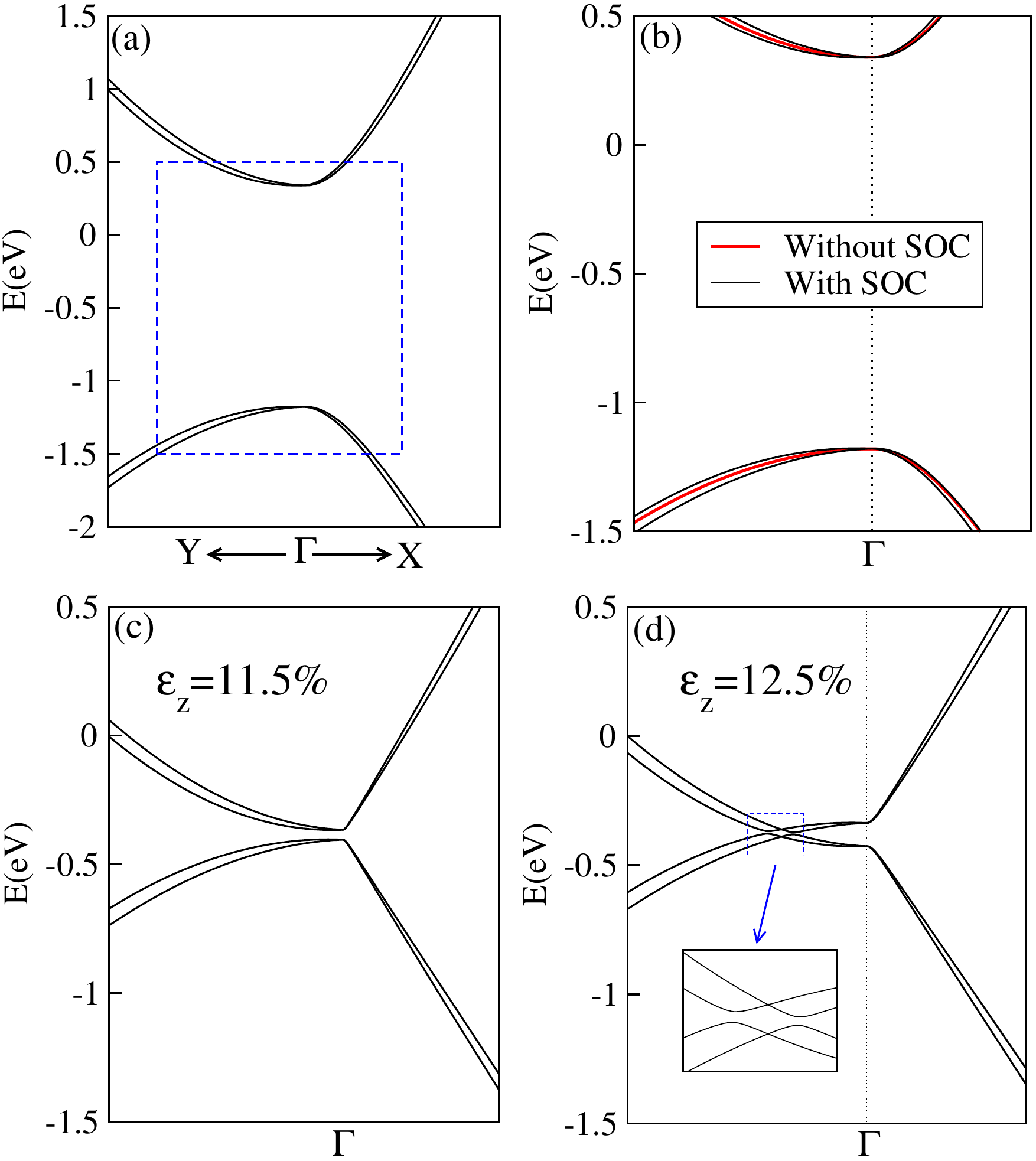}}
 \caption{(a) The TB bands of phosphorene including the effect of SOC. The blue dashed rectangle is
 magnified in (b), (c) and (d) for various conditions:
  (b) The magnified valence and conduction bands of Phosphorene. Red curves show the bands without 
 SOC. Black solid curves show the bands with SOC. (c), (d) The energy spectrum right before and after band inversion at 11.5\% and 
12.5\%  perpendicular tensile strain, respectively. The inset shows the gap opening due to  the SOC, i.e. $\sim5$ meV. }
\label{band-inversion}

\end{figure}

 The energy spectrum  of  pristine phosphorene in the absence of strain has been obtained by numerical diagonalization of 
 the TB Hamiltonian Eq.~(\ref{Hk1}) in different symmetry directions as shown in Fig.~\ref{band-inversion}(a). As
 we can see  in Fig.~\ref{band-inversion}(b), the degeneracies of bands have been removed (black lines) slightly 
 due to the SOC in comparison with the case of zero SOC coupling (red lines) 
 except for the time reversal invariant momentas (TRIMs) which are at least doubly degenerate 
 according to the Kramers theorem.

As seen in Fig.~\ref{band-inversion} the gap of phosphorene is located at the $\Gamma$ point which 
is also a TRIM. At this point, the spin up and spin down valence and conduction
bands are degenerate and the change in the gap due to the SOC is very small as compared to the bulk gap. Since 
axial strain doesn't break TRS, the bands at this point remain degenerate. Therefore, when the bulk
gap is modified by an external factor  such as strain, we can safely use the spinless 
Hamiltonian demonstrating the general trend in changes of the gap.
All P atoms in a unit cell have the same on-site energy, so we can project 
the position of upper and lower chains of phosphorene on a horizontal plane to reduce the spinless $4\times4$~
Hamiltonian $H^{(4)}_{\bf k}$ into a two-band TB model~\cite{sisakht,ezawa2}. Therefore the new $k$-space Hamiltonian 
of the strained phosphorene in the absence of spin is given by 

\begin{figure*}
\centering
{\includegraphics[angle=0,width=1\textwidth]{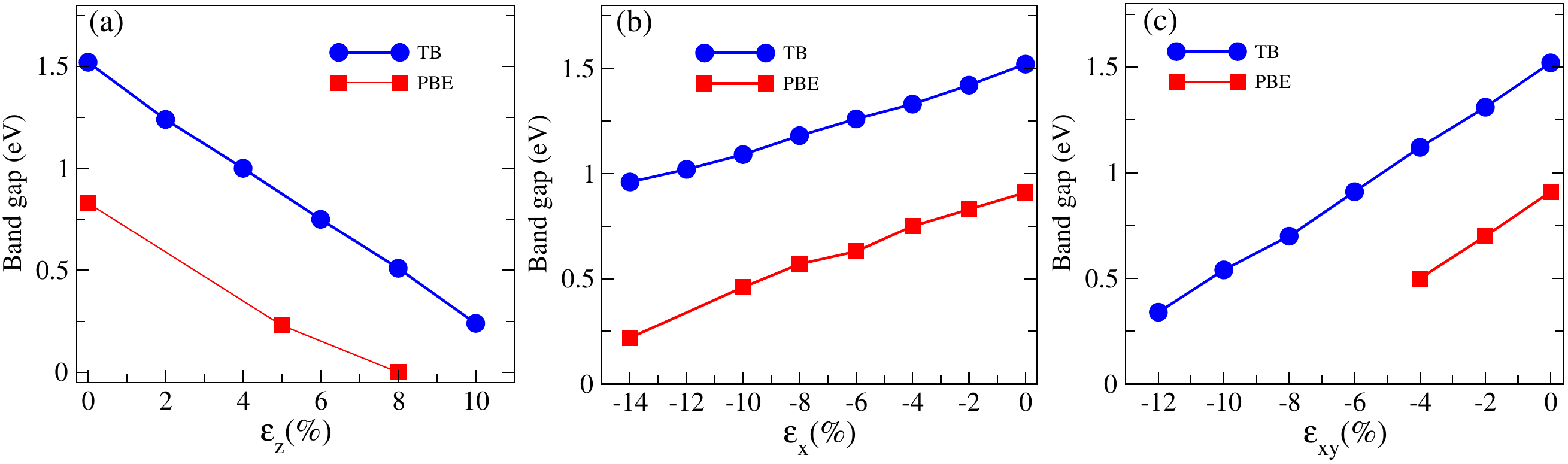}}
 \caption{ Band gap evolution of phosphorene in the presence of (a) perpendicular tensile strain, (b) uniaxial compressive strain in armchair direction, and (c) biaxial compressive in-plane strain. }
\label{gap_ez}
\end{figure*}

\begin{small}
\begin{flalign}
\small H^{(2)}_{\bf k}& = \begin{pmatrix} B_{\bf k}e^{i(k_a-k_b)/2} &\small A_{\bf k}+C_{\bf k}e^{i(k_a-k_b)/2}\\
\small A^{*}_{\bf k}+C^{*}_{\bf k}e^{-i(k_a-k_b)/2}& \small B_{\bf k}e^{i(k_a-k_b)/2}  
\label{Hk2}
\end{pmatrix}.
\end{flalign}
\end{small}
Diagonalizing this Hamiltonian at the $\Gamma$ point gives the band gap as
\begin{small}
\begin{eqnarray}
E_g&=&(4t^0_1+2t^0_2+4t^0_3+2t^0_5)\notag\\
&-&\sum_j(8\alpha^1_j\varepsilon_j t^0_1+4\alpha^2_j\varepsilon_j t^0_2+8\alpha^3_j\varepsilon_j t^0_3+4\alpha^5_j\varepsilon_j t^0_5),
\label{eg0}
\end{eqnarray}
\end{small}where $j$ denotes the summation over $x$, $y$, $z$ components. The first bracket is the unstrained 
band gap i.e. $E^0_g=1.52$~eV and the second one indicates the  structural dependent values of changes in the band gap 
due to the axial strains. Inserting the numerical values of the structural parameters in Eq.~(\ref{eg0}) we obtain a compact form for the gap equation
\begin{eqnarray}
E_g=E^0_g-\sum_j\eta_j\varepsilon_j,
\label{eg}
\end{eqnarray}

where $\eta_x=-4.09$~eV, $\eta_y=-5.72$~eV and $\eta_z=12.86$~eV. Eq.~(\ref{eg}) shows that by applying in-plane compressive 
biaxial strain and perpendicular tensile strain, the band gap decreases which is consistent with DFT calculations~\cite{rodin,peng,wang,zhang,huang}.
It is shown that DFT calculations using the PBE exchange correlation functional anticipate properly the general trends of the band structure when applying axial strains on phosphorene~\cite{peng,wang}.
A comparison between the band gaps as function of axial strains using available DFT data~\cite{peng,wang,huang} and TB model demonstrate that the modification of the hopping parameters in the linear regime 
are valid for rather large strains and show that the modified TB model predicts correctly the variation of the low energy spectrum. Figure~\ref{gap_ez} shows the band gap values evaluated at the $\Gamma$ point in the
presence of (a) uniaxial perpendicular tensile strain (b) uniaxial compressive strain in armchair direction, and (c) biaxial compressive in-plane strain, respectively.
In both DFT and TB approaches the band gaps exhibit linear dependence with applied strain. The discrepancy between the values of the band gaps originate from the specific calculation method.
As a particular case we consider the modification of energy the spectrum under a perpendicular tensile strain.
By increasing the tensile strain, a band inversion occurs at the critical value of $\varepsilon^c_z=E^0_g/\eta_z=0.118$. This 
is a signal of a topological phase transition. Figs.~\ref{band-inversion}(c), (d) show the low energy bands just before and after band closing at 11.5\% and 
12.5\% tensile strain, respectively. As shown in the inset of Fig.~\ref{band-inversion}(d), the SOC opens a 
small gap of about 5~meV after band closing preventing the formation of a Dirac like-cone.

Notice from Figs.~\ref{band-inversion}, that the low energy bands in the armchair direction become more linear under strain. This makes the intra-chain n-n neighbours 
less important justifying the use of the SOC terms of Eq.~(\ref{HSO2}).

\section{TOPOLOGICAL PHASE TRANSITION OF PHOSPHORENE UNDER STRAIN}
The $\mathbb{Z}_2$ classification provides a very strong distinction between two different time  reversal 
topological and trivial phases. Pristine phosphorene as a trivial insulator when the intrinsic SOC
effect is included preserves the TRS and can exhibit a quantum spin Hall (QSH)
phase when its electronic properties is influenced by external factors e.g. electric field or strain.
In the following, we first briefly describe our approach for calculating the $\mathbb{Z}_2$ invariant. This approach, when working in the frame of the TB model~\cite{sol} is quite efficient for 
2D materials such as phosphorene. Then, we will demonstrate numerically  a topological phase transition 
in strained phosphorene and calculate the phase diagrams accordingly. Finally we will show the existence of protected edge states in zPNRs
and discuss their fascinating properties.

\subsection{Calculation of $\mathbb{Z}_2$ invariant}
Fu and Kane~\cite{fu1} showed 
that an equivalent  way to calculate the $\mathbb{Z}_2$ invariant is as an integral over  half the Brillouin zone given by 
\begin{equation}
\small\mathbb{Z}_2=\frac{1}{2\pi i}\left [\oint_{\small{\partial\textrm{HBZ}}}d\bm{k}\bm{\cdot\mathcal{A}}(\bm{k})-
\int_{\small{\textrm{HBZ}}}d^2k \mathcal{F}(\bm{k})]\right]\textrm {(mod 2)},
\label{z2-1}
\end{equation}\\
where  $\small{\textrm{HBZ}}$ denotes half the Brillouin zone. 
$\bm{\mathcal{A}}(\bm{k})=\sum_n\langle u_n(\bm{k})|\nabla_n u_n(\bm{k})\rangle$ is 
the Berry gauge potential and the Berry field strength is written as
$\mathcal{F}=\nabla_{\bf k} \times {\mathcal{A}(\bm{k})\mid_z}$ where $u_n(\bm{k})$ is 
the periodic part of the Bloch state with band index $n$ and  the summation runs over all occupied states.
According to Stoke's theorem,
it is obvious that if $\bm{\mathcal{A}}$ and $ \mathcal{F}$ have the same gauge which is smooth over $\small{\textrm{HBZ}}$, the result will vanish. 
Therefore, one needs to fix the gauge
with some additional constraints~\cite{sol1}. By choosing a gauge, in which the corresponding states fulfills the TRS 
constraints
in addition to the periodicity of the $k$ points, that are related by a reciprocal lattice $\bm G$, the gauge fixing procedure is complete and the returned results of
$\mathbb{Z}_2=0$ or $\mathbb{Z}_2=1$ 
represents the trivial and topological phases, respectively. In the case of  phosphorene,  where bands cross or degeneracies are
present in the  energy spectrum, the Berry potential and Berry field strength must be extended to  non-Abelian gauge field  analogies~\cite{fukui}
associated with a ground state multiplet $|\psi(k)\rangle=(|u_1(k)\rangle,...,|u_{2M}(k)\rangle)$ in the equation $\mathcal{H}(k)|u_n(k)\rangle= E_n(k)|u_n(k)\rangle $.

Based on the above extension, the discretized  Brillouin zone version~\cite{fukui1} of Eq.~(\ref{z2-1}) for numerical computing the $\mathbb{Z}_2$ 
invariant, is written as
\begin{equation}
\small\mathbb{Z}_2=\frac{1}{2 \pi i}\left[ \sum_{{\textit k_l}\in \small{\partial\textrm{HBZ}}}A_x({\textit k_l}) - 
\sum_{{\textit k_l}\in \small\textrm{HBZ}} F_{xy} ({\textit k_l})\right]\textrm {(mod 2)},
\label{z2-2}
\end{equation}
where each site in the square lattice of the Brillouin zone of phosphorene is labeled by  $\textit k_l$ and $\textit l$ specifies 
a plaquette with so-called unimodular link variable
\begin{equation}
U_\mu({\textit k_l})=\frac{\textrm {det}  \psi^\dagger(\textit k_l)\psi(\textit k_l+ \hat{\mu})}{|\textrm {det}  \psi^\dagger(\textit k_l)\psi(\textit k_l+ \hat{\mu})|},
\end{equation}
where $\hat{\mu}$ denotes a unit vector in $x$-$y$ plane. Such a link variable allows us to define the Berry potential and Berry field as
\begin{eqnarray}
 A_{x}({\textit k_l})&=&\ln U_x({\textit k_l}), \\
\small F_{xy}(\textit k_l)&=&\ln\frac{U_x(\textit k_l)U_y(\textit k_l+\hat{x})}{ U_y(\textit k_l)U_x(\textit k_l+\hat{y})}.
\label{FFF}
\end{eqnarray}
Berry potential and Berry field strength are both defined within the branch of $A_{x}({\textit k_l})/i\in(-\pi,\pi)$ and $F_{xy}(\textit k_l)/i\in(-\pi,\pi)$.

\begin{figure}
\centering
{\includegraphics[angle=0,width=0.48\textwidth]{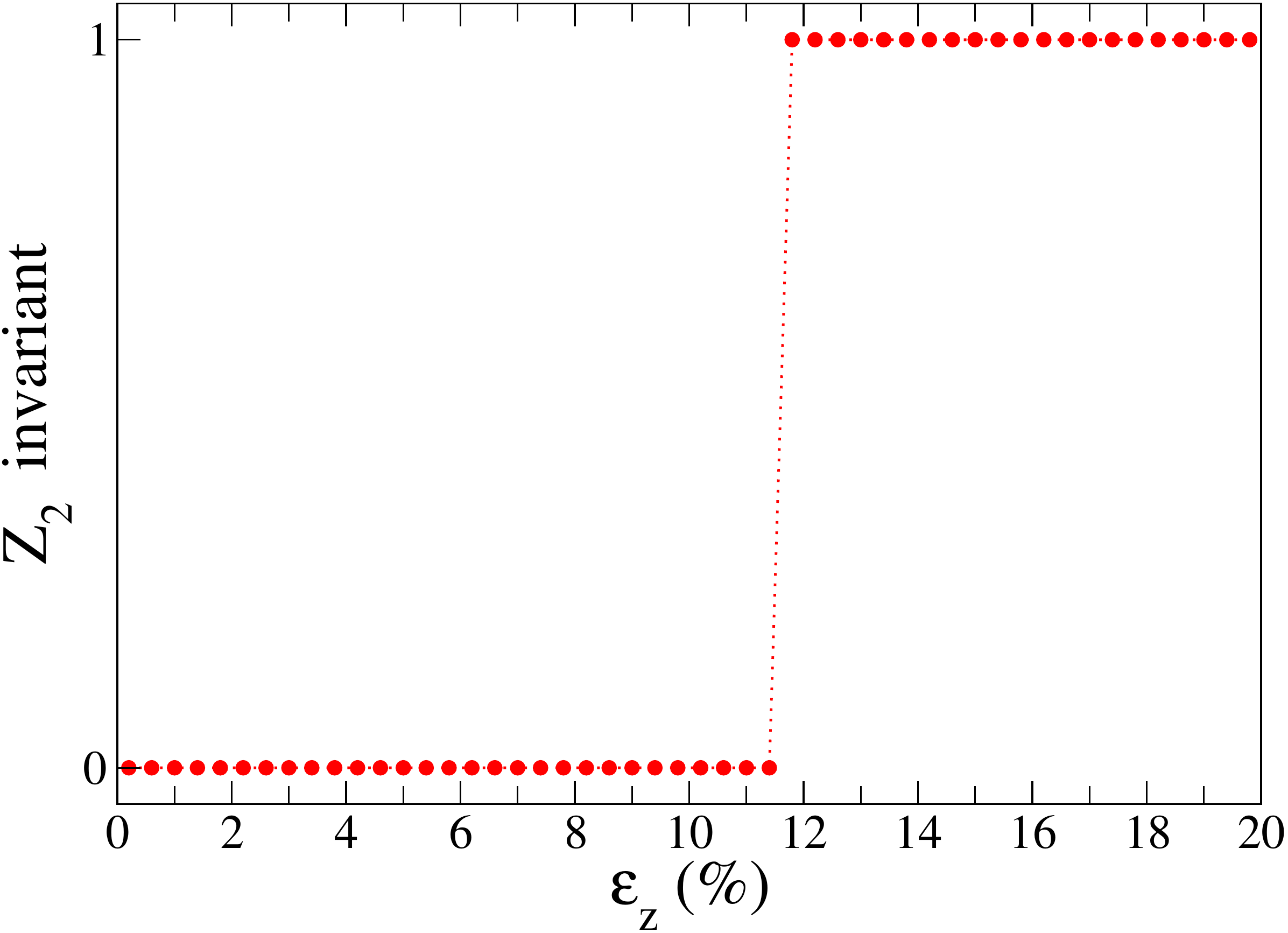}}
 \caption{  Calculation of $\mathbb{Z}_2$ invariant of phosphorene in the presence of perpendicular tensile strain. The critical value for the topological phase transition is $11.8\%$.}
\label{fig4}
\end{figure}

\begin{figure}
\centering
{\includegraphics[angle=0,width=0.47\textwidth]{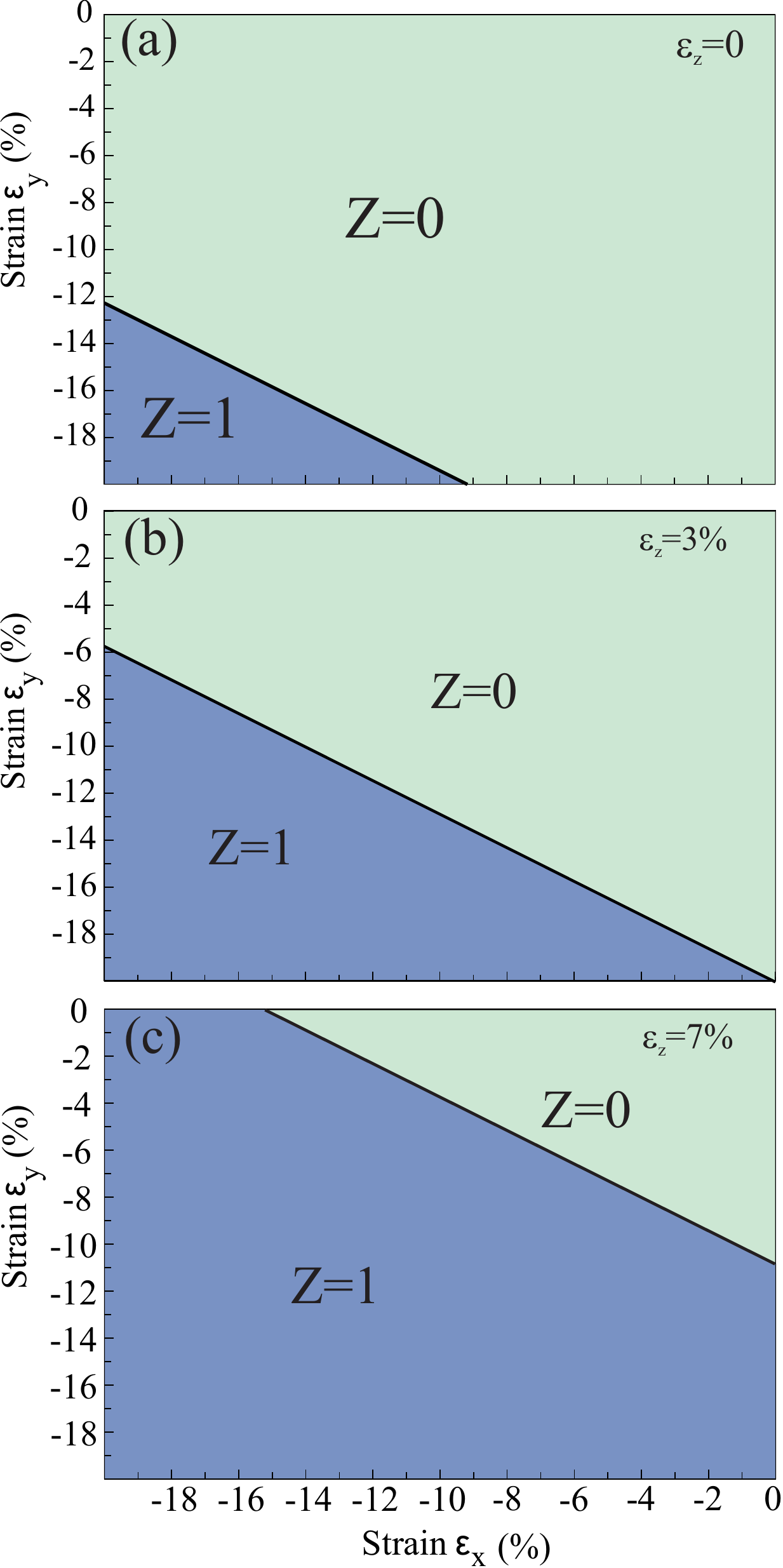}}
 \caption{Phase diagrams of the $\mathbb{Z}_2$ invariant as function of $\varepsilon_x$ and $\varepsilon_y$ for different values of $\varepsilon^c_z$.  
The linear boundaries distinct the two topologically different phases according to the gap closing condition of $\eta_x\varepsilon_x+\eta_y\varepsilon_y=E^0_g-\eta_z\varepsilon^c_z$. }
\label{z-2}
\end{figure}

Figure~\ref{fig4} shows the results of $\mathbb{Z}_2$ corresponding to the energy bands in Fig.~\ref{band-inversion}.
 As can be seen, at the critical strain of $11.8\%$, which is consistent 
 with the condition of $\varepsilon_z>E^0_g/\eta_z$ for band inversion, the $\mathbb{Z}_2$ invariant jumps from $0$ to $1$. This, demonstrates a topological  phase 
 transition in the electronic properties of phosphorene.
 According to  Eq.~(\ref{eg}), another way to observe a topological phase transition in phosphorene, is by applying in-plane compressive biaxial 
 strain at a fixed value of tensile strain in the $z$ direction. Figs.~\ref{z-2} show the  numerically computed $\mathbb{Z}_2$ phase diagrams as a function of $\varepsilon_x$ and $\varepsilon_y$ at a fixed value
 of $\varepsilon_z$.  As can be seen, there is a linear border between two distinct topological phases that corresponds to the regimes before and after the gap closing condition of 
 $\eta_x\varepsilon_x+\eta_y\varepsilon_y=E^0_g-\eta_z\varepsilon^c_z$, where $\varepsilon^c_z$ is a fixed value of strain in the direction of $z$.
 
It is worth mentioning that, the relatively large bulk band gap of monolayer phosphorene necessitates a rather 
large value of strain in order to observe band inversion. As mentioned before, according to DFT calculations, this is accompanied by an upward shift of
a new VBM. After a critical percentage of strain, a direct band touching occurs, which is characterized by a TI phase.  However, 
further increase of strain leads to a metal phase and because the topological 
nature does not change, the system may fall into the TM phase. Our model can not predict the 
VBM upward shift, hence, in spite of demonstrating the change of the topological phase, it can not distinguish between the TI and TM phases. 

Note that our approach can be simply extended to the case of few-layers phosphorene in which we expect to observe the topological phase
transition at lower strain values, due to the fact that the inter-layers hoppings result in a smaller gap~\cite{cai}.

\subsection{Electronic properties of phosphorene nanoribbons under strain}

In this subsection, we investigate the evolution of the band structure of phosphorene nanoribbons in the presence of in-plane and perpendicular strain.
In the following, we refer to the width of zPNRs as $N_z$-zPNR  with $N_z$  being the number
of zigzag chains  across the ribbon width. As we showed
in the previous section, a topological phase transition occurs in the band spectrum of phosphorene. This should lead to the formation of topologically protected edge states in the band structure  
of the corresponding  nanoribbons. 
We obtain the eigenvalues and eigenvectors using the following matrix

\begin{equation}
M_{i\alpha,j\beta}({\bf k}) = \sum_{mn}\tau_{mi\alpha,nj\beta} e^{i{\bf k}\cdot{\bf R}_{mn}},
\label{band1}
\end{equation}
where $e^{i{\bf k}\cdot{\bf R}_{mn}} $ are the 1D Bloch wave functions. $m$,~$n$ denote super-cells; $i$,~$j$ are the basis sites in a super-cell and $\alpha$,~$\beta$ denote 
the spin degree of freedom. ${\bf k}$ is the wave vector, and
${\bf R}_{mn}$ represents a Bravais lattice vector. $\tau_{mi\alpha,nj\beta}$ are
the hopping integrals with  usual SOC  or intrinsic Rashba coupling  that are  conveniently defined  between the basis site $i$ with spin $\alpha$ of super cell $m$
and the basis site $j$  with spin $\beta$ of unit cell $n$.

\begin{figure*}
\includegraphics[angle=0,width=.9\textwidth]{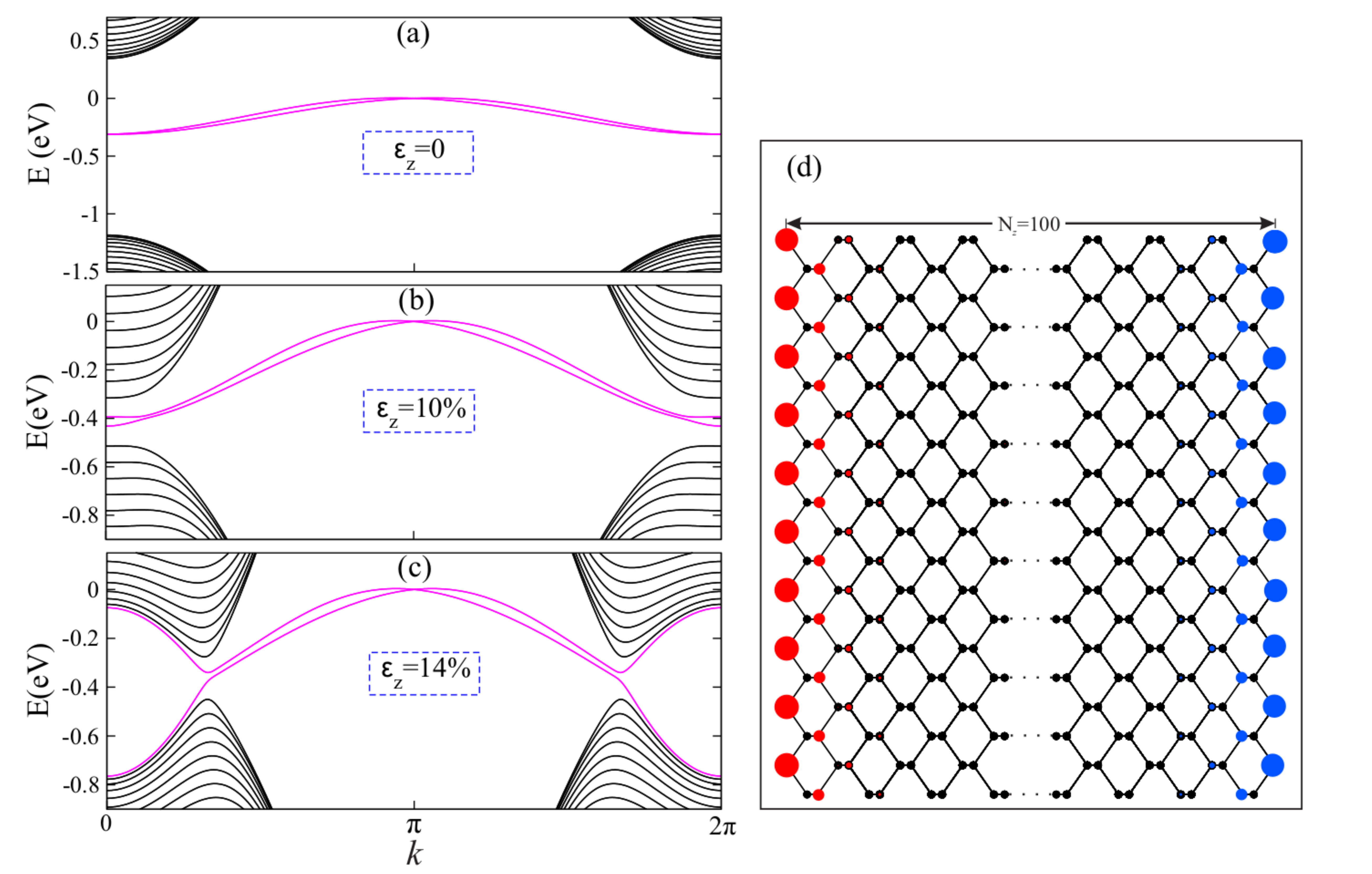}
 \caption{1D energy bands for a typical phosphorene nanoribbon with $N_z=100$~($\sim23$~nm) in case: (a) without strain, (b) $\varepsilon_z=10 \%$, and (c) $\varepsilon_z=14 \%$. (d) The
 amplitude probability of the topological edge modes living on opposite edges for a definite $k$ point. }
\label{edge-modes}
\end{figure*}

Note that, Eq.~(\ref{band1}) is related to the energy spectrum of nanoribbons that are not edge passivated. The experimental realization of such nanoribbons with pristine edges in 
low dimensional materials as graphene is well known~\cite{zhang1} and may be extended to the  case of phosphorene nanoribbons. However, the stability of such ribbons is important 
from the experimental point of view. Formation energy studies~\cite{carvalho} showed that pristine phosphorene nanoribbons are stable specially for ribbon widths which we
have considered in this paper.

The emergence of quasi-flat bands which are detached completely from the bulk bands due to the special structure of phosphorene are well known~\cite{ezawa2,sisakht,gruj}.
As shown in Fig.~\ref{edge-modes}(a), there are topologically non-protected edge modes in the 1D bands of a typical zPNR (the results are for $N_z=100$).
These quasi-flat bands have been used to propose a field-effect transistor driven by an in-plane electric 
field~\cite{ezawa2,sisakht}. However, since pristine bulk phosphorene is a trivial insulator, the existence of topologically non-protected
edge modes in the corresponding nanoribbons which can be affected by environmental conditions such as disorder or 
impurities, may not be a good candidate for practical
use. As an example, we consider the zigzag nanoribbon in the presence of perpendicular strain. 
The behaviour in the presence of other types of strain is similar to this case. As can be seen in Figs.~\ref{edge-modes}(b) and (c), by increasing strain the bulk gap of the
nanoribbon gradually decreases and after a critical strain, where 
a band inversion  occurs in the bulk spectrum, the corresponding edge states in the ribbon cross the gap which demonstrates 
a topological insulator phase. Owing to the 
dependence of the nanoribbon gap on the ribbon width, the critical strain for driving it to a topological insulator phase 
depends on the width as well. If we consider 
ribbons with very large widths, the critical value approaches the critical strain value of bulk 11.8\% that we have calculated in previous section.

The anisotropic structure of phosphorene results in a large bulk gap for zigzag nanoribbons
with experimentally accessible widths. This makes strained zPNRs ideal systems for observing topological states even at room temperature. As shown in Fig.~\ref{edge-modes}(c) 
for a zigzag nanoribbon of width $\sim23$~nm this gap is about 200~meV which is much larger than room temperature  thermal energy. We have
calculated numerically these bulk gaps for relatively large ribbons up to a width of $100$~nm and found that the mentioned gaps are at least three orders of 
magnitude larger than the thermal energy at room temperature.It is worth mentioning that, such a typical ribbon width is wide enough to 
prevent from overlapping of edge states living on opposite sides of the ribbon. The corresponding amplitude probability of the topological edge modes of 
Fig.~\ref{edge-modes}(c) which have amplitude on opposite edges are shown in Fig.~\ref{edge-modes}(d) for a definite $k$ point. The amplitude of the
wave functions drop very quickly along the width of the ribbon demonstrating that the nanoribbon width is wide enough to prevent quantum tunneling.
Such excellent properties
can pave the way for utilizing it in device applications.
\section{Conclusions}
In summary, we derived a spin-orbit model Hamiltonian based on the structural and electronic properties of phosphorene 
that captures the main physical properties of spin-orbit related subjects. Then we showed in the frame of this TB model
that gap engineering of phosphorene by axial strains can lead to a topological phase transition in the
electronic properties of phosphorene. In spite of the relatively small gap induced by SOC in bulk monolayer phosphorene, we predict that due to the special puckered structure 
of phosphorene, zigzag  nanoribbons in the regime of TI have topologically protected edge states with rather large bulk
band gaps of about  $200$~meV for a typical ribbon of width $\sim23 $~nm. Such gaps are larger that the thermal energy at room  temperature and are therefore sufficiently large for practical 
device engineering at room temperature.

\section*{Acknowledgement}
This work was supported by Iran's ministry of science. M.Z. is a postdoc fellow of the Felamish Research Foundation (FWO-Vl).

\bibliography{paper}{}

\newpage

\end{document}